%% file: main.tex
\documentclass[preprint,12pt]{elsarticle}

\usepackage{amsmath,amssymb,amsfonts}
\usepackage{xcolor}
\usepackage{algorithmic}
\usepackage{graphicx}
\usepackage{booktabs}
\usepackage{textcomp}
\usepackage{multirow}
\usepackage{makecell}
\usepackage[hyphens,spaces,obeyspaces]{url}
\usepackage[pdfauthor=author, colorlinks=true, allcolors=blue]{hyperref}
\usepackage[nameinlink,noabbrev]{cleveref}
\usepackage[anythingbreaks]{breakurl}

\journal{Computer Networks}

\begin{document}

\begin{frontmatter}



\title{Website Fingerprinting on Early QUIC Traffic}

\author[1,3]{Pengwei Zhan}
\ead{zhanpengwei@iie.ac.cn}
\address[1]{Institute of Information Engineering, Chinese Academy of Sciences, Beijing, China}
\address[3]{School of Cyber Security, University of Chinese Academy of Sciences, Beijing, China}

\author[1]{Liming Wang}
\ead{wangliming@iie.ac.cn}

\author[2]{Yi Tang\corref{2correspondingauthor}}
\cortext[2correspondingauthor]{Corresponding author}
\address[2]{School of Mathematics and Information Science, Guangzhou University, Guangzhou, China}
\ead{ytang@gzhu.edu.cn}

\begin{abstract}
Cryptographic protocols have been widely used to protect the user's privacy and avoid exposing private information. QUIC (Quick UDP Internet Connections), including the version originally designed by Google (GQUIC) and the version standardized by IETF (IQUIC), as alternatives to the traditional HTTP, demonstrate their unique transmission characteristics: based on UDP for encrypted resource transmitting, accelerating web page rendering. However, existing encrypted transmission schemes based on TCP are vulnerable to website fingerprinting (WFP) attacks, allowing adversaries to infer the users' visited websites by eavesdropping on the transmission channel. Whether GQUIC and IQUIC can effectively resist such attacks is worth investigating. In this paper, we study the vulnerabilities of GQUIC, IQUIC, and HTTPS to WFP attacks from the perspective of traffic analysis. Extensive experiments show that, in the early traffic scenario, GQUIC is the most vulnerable to WFP attacks among GQUIC, IQUIC, and HTTPS, while IQUIC is more vulnerable than HTTPS, but the vulnerability of the three protocols is similar in the normal full traffic scenario. Features transferring analysis shows that most features are transferable between protocols when on normal full traffic scenario, which enable the adversary to use features proven effective on a special protocol efficiently attacking a new protocol. However, combining with the qualitative analysis of latent feature representation, we find that the transferring is inefficient when on early traffic, as GQUIC, IQUIC, and HTTPS show the significantly different magnitude of variation in the traffic distribution on early traffic. By upgrading the one-time WFP attacks to multiple WFP Top-$a$ attacks, we find that the attack accuracy on GQUIC and IQUIC reach 95.4\% and 95.5\%, respectively, with only 40 packets and just using simple features, whereas reach only 60.7\% when on HTTPS. Finally, by conducting the attacks on flawed networks, we also 
demonstrate that the vulnerability of IQUIC is only slightly dependent on the network environment.

\end{abstract}


\end{frontmatter}

\section{Introduction}
\label{Introduction}

TCP (Transmission Control Protocol) and UDP (User Datagram Protocol) are two standard transport-layer protocols in the TCP/IP protocol suite. TCP provides reliable transmission, and packets are transmitted without error, loss, and duplication. Therefore, applications that require reliable connections, e.g., accessing web resources (with HTTP), are implemented base on TCP. However, with the increase of the network environment complexity, problems such as Head-of-line blocking (HOL blocking) and re-transmission ambiguity significantly affect the transmission performance of TCP. QUIC (Quick UDP Internet Connections) is a UDP-based cryptographic protocol with built-in cipher suite and optimized multiplexing\cite{hamilton2016quic,rfc8999,rfc9000}, flow control\cite{google-docs-flow}, and congestion control mechanisms\cite{ietf-tools,rfc9002}, which solves TCP transmission performance shortcomings. After years of development, QUIC, which originally designed by Google (GQUIC), has been standardized by IEFT (IQUIC) and been developed into the new generation of HTTP/3, which can achieves the same or better transmission efficiency as HTTPS (equal to HTTP/2 + TLS + TCP) on most network conditions\cite{megyesi2016quick}.

Meanwhile, confidentiality has become the dominant trend of the Internet. As of May 2021, all of the top 100 non-Google websites support cryptographic protocol, and 97 of them use cryptographic protocol by default\cite{googleTransReport}. Encrypted communication directly causing the traditional methods such as deep packet inspection to fail in sniffing the payload of the packets and thus protecting the content the user visited. GQUIC and IQUIC use different schemes to encrypt data, where IQUIC uses the more secure TLS1.3\cite{rfc9001} and GQUIC uses the informal ``QUIC crypto'' encryption protocol \cite{quiccrypto}. Although the ``QUIC Crypto'' used by GQUIC has been proven to be incapable of resisting active attacks from adversaries while guaranteeing the connection speed of GQUIC \cite{LychevJBN15}, it is unknown if it can safely protect the privacy of users under passive attacks.

Existing cryptographic protocols, including GQUIC, IQUIC, and HTTPS, do not change the working models of the application layer while implementing encrypted communication. Instead, a function suite that can perform encryption is added on the existing transport layer, resulting in that the encrypted traffic still highly correlated with the original website. Specifically, encryption does not significantly change the size of web resources, and the request-response traffic triggered by browser rendering of web pages depends on the structure of the website. It makes Website Fingerprinting (WFP) attacks possible, which try to infer the websites that a user is visiting by learning the correspondence between websites and website fingerprints. Website fingerprint refers to a set of features (e.g., number of packets, packets inter-arrival time) extracted from the traffic between client and server when the user is accessing a website. It should be noted that if the website fingerprint used for WFP attack is from incomplete traffic (e.g., the first 10\% packets of complete traffic), the attack will upgrade to early WFP attack. The early WFP attack will bring higher security risks than a normal WFP attack since it does not need to analyze the entire traffic, and the computation is cheaper, which saving up the time for further operations (e.g., connection blocking).

Furthermore, the rendering paradigm of modern browsers may expands the potential security threats of these cryptographic protocols. When accessing web pages with GQUIC, IQUIC, and HTTPS, the interactions between client and server remains the same, which is always a request-response model, resulting in that the transporting sequence of resources roughly remains the same as transmitted via different protocols and may introduce associations between the traffic of different protocols. Moreover, these undesired associations may extend the vulnerabilities present in one protocol to another, e.g., features proven to be effective in WFP attacks on one protocol can be directly used to attack other protocols efficiently, which we defined as the features transferability.

In this paper, we study the vulnerability of GQUIC, IQUIC, and HTTPS to WFP attack from the perspective of traffic analysis. We construct a comprehensive testbed to collect traffic data without distribution bias and study the WFP attack effectiveness in both the early traffic and normal full traffic scenarios. We also explore the feature importance changes and the feature transferability between the traffic of GQUIC, IQUIC, and HTTPS. Two sets of features are designed, and five machine learning algorithms are selected to comprehensively evaluate the vulnerability of GQUIC, IQUIC, and HTTPS. 

Our experiments show that GQUIC and IQUIC, especially GQUIC, are more vulnerable to WFP attacks than HTTPS in the early traffic scenario but are similar in the normal traffic scenario. Meanwhile, features are transferable between protocols, and the feature importance is almost inherited on the normal traffic, while the transferring is inefficient when on early traffic due to the different magnitude of variation in the traffic distribution of GQUIC, IQUIC, and HTTPS. Finally, we conduct the attacks under different network conditions, showing that various imperfections in the natural network environment are only slightly affect the vulnerability of protocols under WFP attacks, and the conclusions we obtained about the vulnerability of GQUIC, IQUIC, and HTTPS can be generalized to the natural network environment.

The main contributions of this paper are as follows:

\begin{itemize}
\item We implement WFP attacks on GQUIC, IQUIC, and HTTPS, demonstrating that GQUIC and IQUIC are more vulnerable to WFP attacks than HTTPS in the early traffic scenario but are similar in the normal full traffic scenario. Even computationally cheap and simple features are highly effective when attacking GQUIC and IQUIC, especially GQUIC.
\item We perform transfer studies among GQUIC, IQUIC, and HTTPS, showing that, when on normal traffic, most features are transferable between protocols, and the feature importance is almost inherited; however, the transferring is inefficient when on early traffic due to the different magnitude of variation in the traffic distribution.
\item We quantitatively analyze the latent feature representation space of the traffic of GQUIC, IQUIC, and HTTPS, finding that features can represent inter-class and intra-class samples more efficiently on GQUIC and IQUIC than on HTTPS, intuitively showing why GQUIC and IQUIC are more vulnerable than HTTPS.
\item We implement upgraded multiple WFP Top-$a$ attacks on GQUIC, IQUIC, and HTTPS, exposing the high insecurity of GQUIC and IQUIC: with only 40 packets, attack accuracy reaches 95.4\% when on GQUIC, reaches 95.5\% when on IQUIC, whereas only 60.7\% when on HTTPS.
\item We implement WFP attacks on IQUIC under different network conditions, showing that bandwidth and delay hardly affect the efficiency of WFP attacks, but a higher loss rate will lead to less effective attacks. However, informative features can resist the uncertainty of network imperfection and the trend of attack effectiveness is always consistent, indicating that the vulnerability of IQUIC is only slightly dependent on the network condition.

\end{itemize}

The rest of this paper is organized as follows. In \Cref{RelatedWorks}, we discuss related work. We introduce QUIC, website fingerprint, and WFP attacks in \Cref{Preliminaries}. In \Cref{DataCollection}, we detail our testbed and how we collect data. We detail our designed features in \Cref{FeatureExtraction}. In \Cref{Evaluation}, we provide the experiment results and related discussion. In \Cref{Discussion}, We discuss the limitations of the experiment and future work. In \Cref{Conclusion}, we conclude this paper.

\section{Related Works}
\label{RelatedWorks}

In this section, we provide an overview of related work on WFP attacks, which is divided into WFP on unencrypted traffic and WFP on encrypted traffic based on the protocol studied. Besides, some critical works on QUIC was discussed.

\textbf{Website fingerprinting on unencrypted traffic.} Karagiannis et al.\cite{karagiannis2005blinc} proposed the BLINC algorithm that analyzed host behavior and successfully identified eDonkey, MSN, IRC, NNTP, and SSH traffic. Bar-Yanai et al.\cite{bar2010realtime} identified the traffic of Skype, eDonkey, and BitTorrent through a hybrid algorithm of k-means and KNN, which analyzed the flow-based features of different applications and achieved approximately 99.1\% accuracy. Lakhina et al.\cite{lakhina2005mining} adopted an entropy-based method and clustering method to distinguish regular traffic from abnormal traffic. Gu et al.\cite{gu2005detecting} divided packets into 2348 categories according to functions and destination port numbers and then performed maximum entropy analysis on the packet classes distribution. Furthermore, a behavior-based model to detect traffic anomalies was established, reaching $F_1$ above 0.93.

\textbf{Website fingerprinting on encrypted traffic.} As early as 1996, Cheng et al.\cite{cheng1998traffic} have researched encrypted traffic analysis attacks. They used the HTML file size and requested traffic size to identify websites. Liberatore and Levine\cite{liberatore2006inferring} used NaiveBayes and Jaccard's coefficient to attack SSH on a dataset containing 2,000 web pages with a total of 480000 samples. Bernaille et al.\cite{bernaille2007early} adopted a method based on GMM clustering, using the packet size to identify SSL connection. Khakpour et al.\cite{khakpour2012information} used entropy vector estimation, Decision Tree, Support Vector Machine to identify encrypted traffic. Zhang et al.\cite{zhang2012encrypted} used an improved k-means algorithm to distinguish between SSH, SSL, and non-encrypted applications. Panchenko et al.\cite{panchenko2016website} proposed the CUMUL method, which utilized 104 features to conduct WFP attacks on the Tor and has achieved more than 91\% accuracy on several datasets, including ALEXA100. Dusi et al.\cite{dusi2009using} used GMM and SVM to identify applications under SSH based on packet size and direction. Tong et al.\cite{tong2018novel} used Random Forest and Convolutional Neural Networks to identify applications under QUIC, considering Netflow-based and Packet-based features, including the size and direction of packets. Five types of Google app services were distinguished with 99.24\% accuracy. Hayes and Danezis\cite{hayes2016k} proposed a Random Forset based WFP model called k-fingerprinting. They used features such as the number of packets statistics, incoming and outgoing packet ratios, and packet ordering statistics to identify 30 monitored services on 100,000 web pages, achieved 85\% true positive rate and 0.02\% false positive rate.

\textbf{QUIC.} Biswal and Gnawali\cite{biswal2016does} showed that QUIC performs better than HTTP/2 in weak network environments if the website does not contain many small-sized objects. Megyesi et al.\cite{megyesi2016quick} compared the performance of HTTP, SPDY, and QUIC in different network situations, finding that QUIC has better performance under the network with slight packet loss and large Round-Trip Time (RTT). They concluded a Decision Tree to reach the best perform protocol in different network conditions. Cook et al.\cite{cook2017quic} found that QUIC outperforms HTTP/S in unstable networks (e.g., wireless mobile networks) but perform similarly in stable and reliable networks. These studies demonstrate the importance of QUIC in improving transmission efficiency.
\section{Preliminaries}
\label{Preliminaries}

\subsection{QUIC Protocol}
There are two versions of QUIC, including the one designed by Google (GQUIC) and released the first version in August 2013, and a new version of QUIC that standardized by IETF (IQUIC) in May 2021 based on GQUIC. GQUIC has made huge changes in design concept\cite{hamilton2016quic}, packet loss recovery and congestion control mechanisms\cite{ietf-tools}, encryption details\cite{langley2016quic}, and flow control mechanisms\cite{google-docs-flow} to address the performance bottlenecks of TCP. GQUIC only requires userspace support, making it possible to deployed easily without changing the Internet middleware\cite{hamilton2016quic}. About 5.5\% of websites support GQUIC, including Youtube and Google Translate\cite{w3techs}. GQUIC is a protocol that uses UDP to send HTTP/2 frames, while IQUIC is a general transport protocol, i.e., protocols other than HTTP can also be transported over IQUIC. Specifically, the HTTP mapping over IQUIC is called HTTP/3, which is supported by 19.8\% of websites. In addition to further optimizing the performance\cite{rfc9000,rfc9002,rfc8999} of IQUIC compared to GQUIC, IQUIC also uses TLS1.3\cite{rfc9001} to completely replace the informal ``QUIC Crypto'' encryption scheme\cite{quiccrypto} used in GQUIC, making it more secure.

\subsection{Website fingerprinting}
A website is a specified webpage $W$ which can be viewed as a Web resource set $R = \{r_i\}$. To render this page, a user agent, in general a browser, will fetch those resources and generate web traffic $T$, $T = \text{Traffic}(R) = \{(t_j, d_j)\}$, where $t_j$ is the $j$-th packet and $d_j$ is the direction of client-to-server (positive) or server-to-client (negative). A feature $F$ is an $m$-dimension vector, $(f_1, f_2, ..., f_m)$, that characterizes the traffic $T$, and is denoted by $F = \text{Feature} (T)$. Ideally, the traffic fingerprinting of website $W$ can be represented by $F$ in a singe visit since the visit traffic $T$ coming from $W$. However, dynamic network communications and parallel resource downloads make the same website generate different traffic in different visits and the feature from a single visit cannot characterize the webpage traffic features. It needs an algorithm to extract a more general feature from traffic in $n$ different visits. When no ambiguity is possible, we reuse the notation $F$ and have $F = \text{Feature} (T_1, T_2, ..., T_n)$, which is called the fingerprinting of website $W$.

Define $\text{Early}(T, k)$ as a function that fetch the first $k$ packets from $T$, i.e., $\text{Early}(T, k) = \cup_{j=1}^{k}\{(t_j, d_j)\}$. It is obvious that $\text{Early}(T, k)$ is in fact the early traffic of website $W$. Following the notation in previous paragraph, we have a feature $\text{Early}(F, k)$ which is also an $m$-dimension vector that characterizes the early traffic $\text{Early}(T, k)$ and also an early fingerprinting of website $W$, $\text{Early}(F, k)$, comes from the early traffic in $n$ different visits.

\subsection{Website Fingerprinting Attacks}
Suppose an adversary wants to infer a specified website visiting by users. The adversary can reside in intermediate nodes such as the routers or the switches to eavesdrop on the network traffic. Since the web page is transported in encrypted over GQUIC, IQUIC, or HTTPS, the adversary cannot identify what content is hidden in traffic because of the limited decryption ability. If an adversary wants to identify the web page a user is visiting from encrypted traffic, the most common way is to find the unencrypted packets (e.g, Handshake packets) in the encrypted traffic and correspond the IP addresses or Server Name Indications (SNI) contained to the real servers. However, the widely used tunnels, proxies, secure gateways, Domain fronting techniques, and Content Delivery Network (CDN) make the IP address and SNI in packets not associated with the actual server, making this traditional identification method fail. Thus, it seems that the privacy of users' visiting behaviors could be well-preserved when visiting over cryptographic protocols. 

However, GQUIC, IQUIC, and HTTPS, the primary application secure protocols for website visiting, are all based on the request-response model, which implies that it is possible to distinguish traffic from different web resources due to the relatively fixed browser rendering sequence. Furthermore, the encryption of web resources is on block cipher, making the sizes of encrypted resources not changed significantly compared to the plain resources. And thus, WFP attacks could be launched.

Suppose there are $N$ websites the user may visit, the adversary needs to continuously maintain a fingerprint series $S=\cup_{i=1}^{N}\{F_i\}$ or $S=\cup_{i=1}^{N}\{\text{Early}(F_i, k)\}$ depended on normal or early traffic scenario on the eavesdropping process. The purpose of the adversary is to learn the parameters $\theta$ of the attack model $M$ from $S$. Then a normal WFP attack and a WFP attack on early traffic can each be represented as:

\begin{equation}
M(F,\theta) \longrightarrow W
\label{WFPattack}
\end{equation}

\begin{equation}
M(\text{Early}(F, k),\theta) \longrightarrow W
\label{EWFPattack}
\end{equation}

According to the definition we have given, the specific WFP attack process is divided into two stages(\Cref{f1}):

\begin{figure}[ht]
\centering
\includegraphics[width=\textwidth]{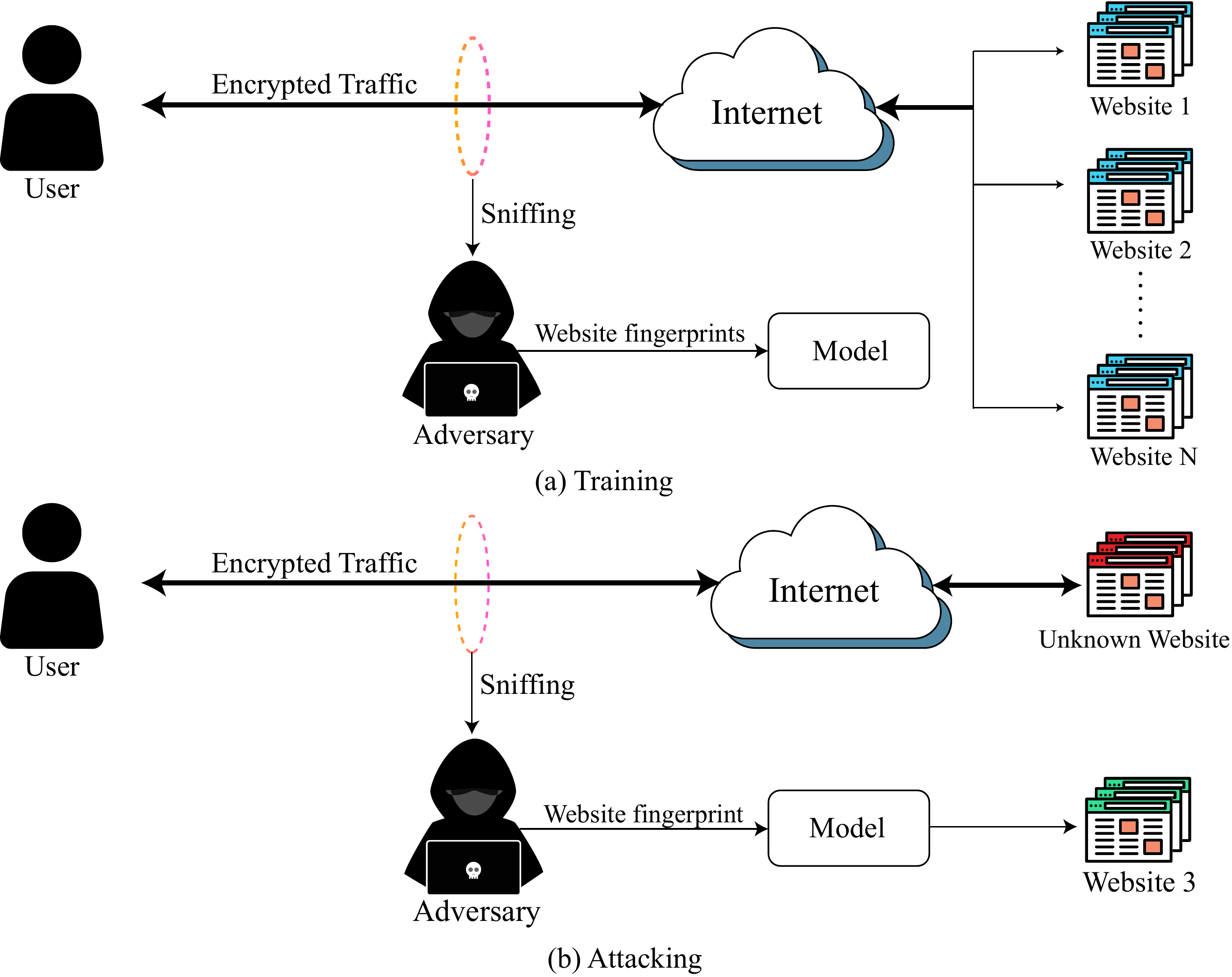}
\caption{Processes of website fingerprinting attack}
\label{f1}
\end{figure}

\begin{enumerate}
    \item \textbf{Training}. Adversary passively eavesdrops on the user's exit-side network traffic, from which website fingerprints are generated and utilized to optimize the attack model (\Cref{f1}(a)). 
    \item \textbf{Attacking}. Adversary generates website fingerprint from specific traffic and predicts the related website by asking the well-trained attack model(\Cref{f1}(b)). 
\end{enumerate}

Even if the plain content that can help identifying the website cannot be sniffed directly from the traffic, the attack model and website fingerprint will still expose the website, causing the privacy benefited from encryption damaged.

\subsection{Website Fingerprinting Attack on Early Traffic}

The browser engine determines the display of web page, and the same page may be displayed differently on different browsers. However, although modern browsers may use different kernels (e.g., Google Chrome: Blink; Mozilla Firefox: Gecko; Safari: WebKit; Internet Explorer: Trident), they all follow a similar paradigm to fetch web resources when rendering a certain web page, as the following steps:

\begin{enumerate}
\item \textbf{Requiring basic HTML file}. As the user starts visiting a web page, the browser synchronously sends a request to the server, and fetches the basic HTML file.

\item \textbf{Requiring web resource files}. As the browser starts rendering the basic HTML file, a sequence of web resources embed in the HTML file may be needed from servers. These resources may include but are not limited to the CSS files, external JavaScript files, image files, etc. So a sequence of web resources requests are sent and the web page visiting traffic is generated.  

\end{enumerate}

It should be noted that different HTML files will cause different web traffic because of different sequences of web resource requests, and the more web resource requests, the more differences will be in traffic. A regular WFP attack uses all traffic packets to build the website fingerprint (from start to end of entire traffic), but an attack on early traffic, instead, use packets from the incomplete early traffic that in the front of the entire traffic (from first to $k$-th packet of traffic).

The effect of WFP attacks on early traffic is getting closer to normal WFP attacks as the parameter $k$ increases. Intuitively, an attack on early traffic can be performed more quickly than a normal attack since it does not need to wait for the total traffic to be transfer, and computation is cheaper as fewer packets are considered. The WFP attacks on early traffic should complete when the website visited by the user is just finished initializing, enabling the secure gateway or adversary to perform further operations, e.g., implement the RST attacks for TCP connections or blocking the client IP for continue visiting, making the connection between client and server blocked without the user's awareness when the targeted website is visited. Therefore, a WFP attack on early traffic is more threatening than a normal WFP attack.

\subsection{Assumptions}
We refine the assumptions made in the previous section, which ensure a fair comparison of GQUIC, IQUIC, and HTTPS and guarantee that extraneous factors do not confound our conclusions on vulnerability and feature transferability.

\begin{itemize}
\item \textbf{Closed-world}: The page a user may visit is limited to $N$ pages. This assumption reduces the size of $S$ that the adversary needs to maintain, ensuring that every $T$ generated by the user can be collected to update the maintained $S$, with no additional filtering required by the adversary. Conversely, users can access pages other than the $ N $ target pages in the open-world scenario.
\item \textbf{Traffic parsing ability}: An adversary can eavesdrop on the user to collect traffic on the user's exit-side and detect the start position and end position of traffic. However, the end position of traffic is not needed to identify on early traffic scenario but the $k$-th packet position.
\item \textbf{Normal computing power}: An adversary can complete the construction of the website fingerprint within an acceptable time delay, and cannot decrypt or modify the packet.
\item \textbf{Same-origin resource limit}: All target pages only contain same-origin resources, including CSS scripts, JavaScript scripts, images and video resources, etc.

\end{itemize}

\section{Data Collection}
\label{DataCollection}

To explore the vulnerability of GQUIC, IQUIC, and HTTPS under the WFP attacks and the transferability of features, we require a dataset consisting of encrypted traffic that can eliminates the impact of data distribution differences and simulates the condition that a real user is visiting web servers. The dataset must (i) be large enough to indicate the pattern in real world (dynamic network communications and parallel resource downloads happens), (ii) contains aligned encrypted traffic of GQUIC, IQUIC, and HTTPS, which originated from the same group of resources, and (iii) be collected like what an actual adversary would attempt. In this section, we present the details of our testbed and the processes we collect data.

\subsection{Testbed Setup}
\begin{figure}[ht]
\centering
\includegraphics[width=\textwidth]{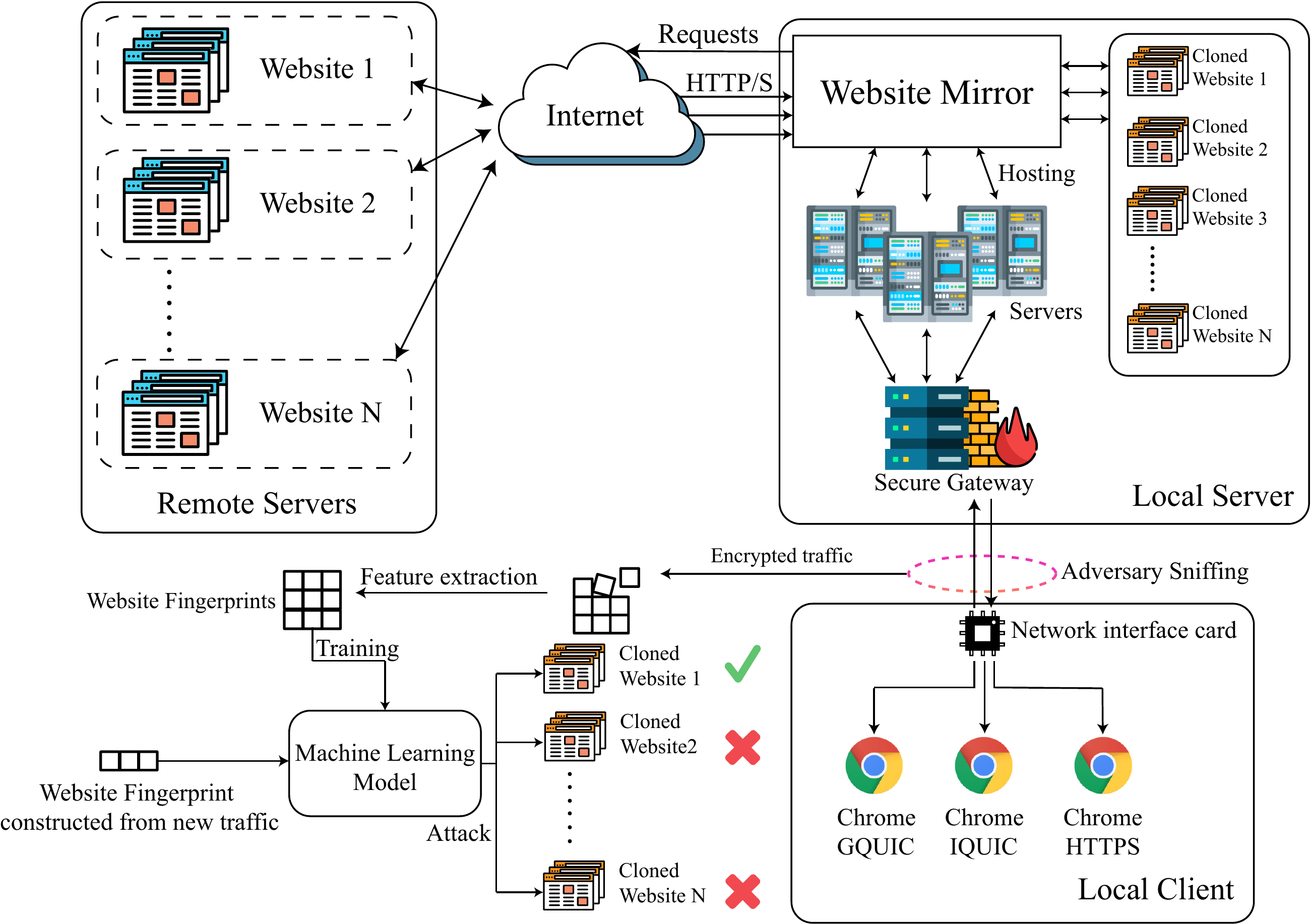}
\caption{Architecture of established testbed}
\label{f2}
\end{figure}

To collect qualified traffic data, we establish a comprehensive testbed. The overall architecture of the testbed is shown in \Cref{f2}.

The entire testbed is established under a controlled environment, and we do not sniff traffic data from the Internet directly. Instead, we sniff from a controlled network for (i) it can prevents individual functional packets (e.g., RST packets in TCP) from disturbing the regular traffic, and ultimately ensuring the reliability of experiment results and (ii) only a few websites support both GQUIC, IQUIC, and HTTPS, and the complete details of these sites are difficult to obtain. We expect to collect aligned GQUIC, IQUIC, and HTTPS traffic originating from the same group of resources, avoiding the distribution differences of the traffic of three protocols. 

We select the official landing pages of the top 100 schools in the 2019 TIMES World University Rankings \cite{timeshighereducation} as our closed-world. We first utilize HTTrack Website Copier\footnote{\url{https://www.httrack.com/}} to clone the same-origin resources of the target website (e.g., A.com/figure.jpg is the same-origin resource of A.com, while img.A.com/figure.jpg is not.) that hosting on the remote servers with parameter depth set as 2 (i.e., traverse all the files in the first and second levels directory of the root directory) to the local Server, a machine with Ubuntu 18.04 installed. Cloned website resources include HTML files, CSS files, JavaScript files, pictures, and other media files involved in the entire process of rendering the landing page.

We utilize three machines, with two Intel(R) Xeon(R) CPU E5-2680 v4 @ 2.40GHz, 128GB RAM, and system Ubuntu 18.04, as web hosting servers. On each server, we utilize Docker to further isolate CPU, memory, and disk resources in containers so that the websites deployed on one machine would not affect each other. All websites are deployed split between three servers, with one machine deploying 30 websites and the other two deploying 31 websites each ($N=92$, detailed in \cref{CollectedDataandTraffic}). We run different versions of Caddy Server\footnote{\url{https://caddyserver.com/}} in Docker to ensure that GQUIC, IQUIC, and HTTPS are properly supported. For collecting GQUIC and HTTPS traffic, we use a Caddy Server in version v1.03, and use a Caddy Server in version v2.4.1 when collecting IQUIC traffic. The GQUIC is in version Q043, the IQUIC is in version h3-29, and the HTTPS is on HTTP/2. Specifically, we set up a Secure Gateway on the edge of the Local Server part as a reverse proxy server to forward all client-server connections to ensure that adversaries cannot get information about the real server from the packets. The Secure Gateway is a machine with a fixed IP address running Ubuntu 18.04, implementing reverse proxy with Caddy. In addition, this Secure Gateway can also quickly update the local forwarding policy based on the results of WFP attacks to block access from a client to a specific website. Another host in the same local area network is taken as the client, on which we utilize Selenium\footnote{\url{https://chromedriver.chromium.org/}} to drive Chrome automatically to imitate actual user behavior when visiting a website. To visit the website with GQUIC and HTTPS, we use Chrome in version 71, and Chrome in version 91 when visiting the website with IQUIC. Expressly, Chrome is set to disable cache mode to prevent interference from previous access, and enable QUIC mode is set when website transfer via GQUIC and IQUIC. Specifically, a complete visit process has the following steps:

\begin{enumerate}
\item Selenium starts a new Chrome process;
\item Chrome accesses the target landing page and request all associated resources;
\item Chrome waits for the landing page to render completely and return the success state to Selenium;
\item Selenium waits extra 5 seconds after rendering complete, and then close the current Chrome process;
\end{enumerate}

Although it is not required that the traffic is complete when performing an early WFP attack, we still take measures (e.g., wait for additional time, stop when the web page rendering is complete) to ensure that the traffic is complete, thus better simulating the behavior of the user. We define 100 consecutive visits as a sniffing cycle, and visit each website via GQUIC, IQUIC, and HTTPS, respectively, for 100 times, and utilizing tshark\footnote{\url{https://www.wireshark.org/}} to sniff the traffic on the exit-side of the client's network interface card to simulate the behavior of the adversary(\Cref{f1}). Collected traffic data is in \emph{pcap} format, and each file contains all traffic obtained in a sniffing cycle.

\subsection{Collected Data and Traffic}
\label{CollectedDataandTraffic}
A total of 7.07GB website are cloned, with 92 websites are available after deleting invalid and error pages; therefore, $ N $ in our experiments is 92. \Cref{t1} shows basic information about the three smallest and largest websites, where AVG represent the arithmetic average, $Q_1$ and $Q_3$ indicate the first and third quartiles. It should be noted that the files cloned to the local servers are a superset of the files involved when accessing the landing page, which ensures that the files requested by the client are complete, thus ensuring a better representation of the actual websites.

\begin{table}[ht]
\footnotesize
\centering
\caption{Statistics of collected websites}
\label{t1}
\resizebox{\textwidth}{!}{%
\begin{tabular}{@{}rrrrrr@{}}
\toprule
\multicolumn{1}{r}{URL} & Size & \# of files & File size (AVG) & File size ($Q_1$) & File size ($Q_3$)\\ \midrule
www.usc.edu & 1.9 MB & 88 & 37.6 KB & 4.2 KB & 39.7 KB\\
www.wisc.edu & 2.1 MB & 62 & 54.3 KB & 7.6 KB & 62.9 KB \\
www.unimelb.edu.au & 2.5 MB & 97 & 44.2 KB & 6.9 KB & 66.1 KB\\
$\cdots$ & $\cdots$ & $\cdots$ & $\cdots$ & $\cdots$ & $\cdots$\\
www.tum.de & 458.2 MB & 2502 & 320.7 KB & 51.9 KB & 159.5 KB \\
www.monash.edu & 1592.1 MB & 1661 & 1161.5 KB & 1078.6 KB & 1105.9 KB \\
www.skku.edu & 2536.2 MB & 9977 & 251.4 KB & 23.4 KB & 101.6 KB \\ \bottomrule
\end{tabular}
}
\end{table}

The traffic of each website obtained in a sniffing cycle ranges from 4.5 MB to 1.86 GB (i.e., 45 KB to 18.6 MB for a single visit), and the final traffic data size is 66.41 GB, of which 22.36 GB is from GQUIC, 21.56 GB is from IQUIC, and 22.49 GB is from HTTPS. To explore the vulnerability of protocols at different degrees of early, we define 40 different parameters $k$, and finally, 1,104,000 traffic are extracted in total for three protocols.

\subsection{Traffic Tailoring}
\begin{figure}[ht]
\centering
\includegraphics[width=\textwidth]{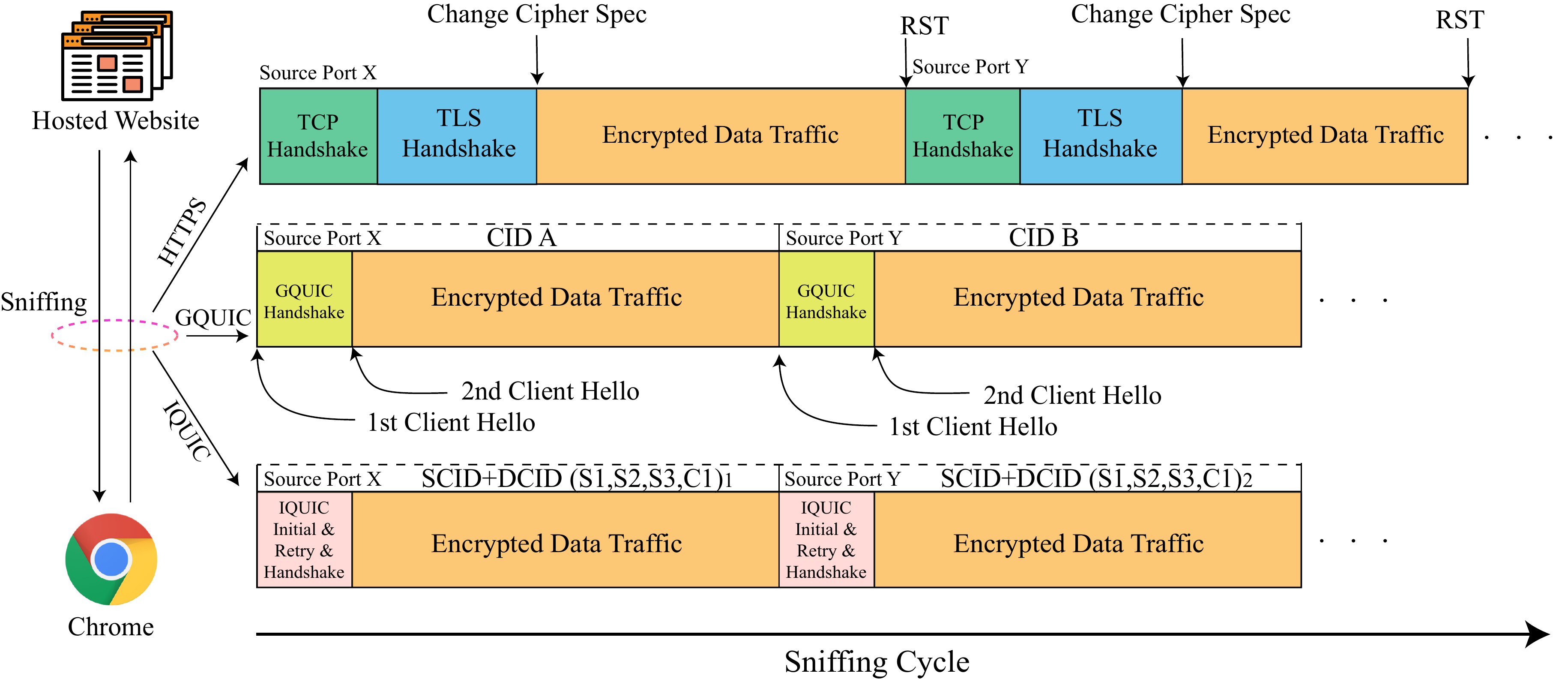}
\caption{Illustration of traffic tailoring}
\label{f3}
\end{figure}

Chrome in ``disable cache'' mode drove by Selenium, will start a new process on each visit, making the client and server always establish a connection as the first-time connection. However, the handshake process is different for GQUIC, IQUIC, and HTTPS during the initial connection establishment, where GQUIC and IQUIC require 1-RTT to complete the handshake, while HTTPS requires 3-RTT. For GQUIC, the client will first generate the handshake parameters with a new Connection ID (CID) and send them to the server in a \emph{Client Hello} packet, which represents the beginning of a GQUIC Handshake. At the end of the GQUIC handshake, the client sends the second \emph{Client Hello} packet. For IQUIC, the complete connection establishment process includes Initial, Retry, and Handshake processes. Multiple Destination Connection ID (DCID) and Source Connection ID (SCID) are used throughout the process to identify the connection, with the client using one (C1) and the server using three (S1, S2, S3). The first-time handshake process of HTTPS is divided into TCP Handshake and TLS Handshake, and a \emph{Change Cipher Spec} packet from the client indicates the end of the TLS handshake and the complete HTTPS handshake.To avoid influence caused by the different handshake processes, we tailor traffic from the first packet after the handshake to the last packet of the current conversation (\Cref{f3}), i.e., traffic used to generate website fingerprint is limited to the \emph{Encrypted Data Traffic}. We utilize \emph{Source Port} to extract different sessions, and further sorting and tailoring the traffic in the sniffing cycle more precisely according to the characteristics of different protocols: in GQUIC, CID indicates different conversations; in IQUIC, a group of DCID and SCID indicate different conversations; in HTTPS, the RST packet close the current conversation.

\section{Feature Extraction}
\label{FeatureExtraction}

The effectiveness of the WFP attacks is directly related to the extracted features, and it is significant to discover informative and straightforward features, which help improving attack efficiency and effectiveness. A common way to discover effective features on a new protocol is to try to use the features that have been proven effective on a previous protocol. Due to the relatively fixed browser rendering processes and the unchanged request-response conversation model, the knowledge represents by the features may share between protocols, which we defined as \emph{feature transferability}. However, the differences between protocols still introduce uncertainty in the features transferring, and the knowledge discovered on HTTPS may be incapable of improving the attacks on GQUIC and IQUIC. To explore the vulnerability of GQUIC, IQUIC, and HTTPS and the transferability of features under both early and normal traffic scenario, we design two groups of features, including the simple and computationally cheap \emph{Simple} features and \emph{Transfer} features that are selected from previous works and are effective in the attack on HTTPS. In this section, we detail the designed features.

\subsection{Simple Features}
The WFP attacks on early traffic can infer the website that the user is visiting much faster than regular WFP attacks. However, the efficiency of WFP attacks, especially in the early traffic scenario, will decrease if the features are complicated and computationally expensive. To efficiently perform attacks on the high speed steam, and better demonstrate the vulnerability of protocols, we design a group of simple but informative features, which are only related to the packet's direction and size, defined as Simple features.

First, we divide packets into two categories based on their direction $d$:

\begin{itemize}
\item Positive: packets from the client to the server.
\item Negative: packets from the server to the client.
\end{itemize}

Second, we divide the packets into four categories based on their size:

\begin{itemize}
\item Tiny: packets size less than 79 bytes.
\item Small: packets size range from 80 – 159 bytes.
\item Medium: packets size range from 160 – 1279 bytes.
\item Large: packets size over 1280 bytes.
\end{itemize}

We divide all packets into eight categories, considering their direction and size: \emph{Positive Tiny, Negative Tiny, Positive Small, Negative Small, Positive Medium, Negative Medium, Positive Large}, and \emph{Negative Large}, and the number of packets in each category constitute the Simple features. Specifically, $T$ can be characterized with $F=\text{Feature}(T)=(n_{pt},n_{nt},n_{ps},n_{ns},n_{pm},n_{nm},n_{pl},n_{nl})$, where $n$ represents the number of different kind of packets in the eight categories.

When a packet is collected, it can be classified into the proper category immediately by just reading the packet header's information, and the computation all is focus on packet counting, which meets our expected low-cost computing characteristics. In our experiments, we also found that although this set of features is simple, it contains a wealth of information in the WFP attacks. This set of features is also a set of controls relative to the intricate features presented subsequently to illustrate that the vulnerability of the protocol is not caused by the defects of specific features.

\subsection{Transfer Features}
In previous studies, many features have been proven effective in traffic identification and WFP attacks on TCP-based protocols, such as unique packet size\cite{liberatore2006inferring}, packet size count\cite{herrmann2009website}, and packet order\cite{ crotti2007traffic}. Features used on HTTPS can be divided into five levels: Packet-level, Burst-level, TCP-level, Port-level, and IP-level\cite{yan2018feature}. Some effective features are selected among Packet-level and Burst-level, since TCP-level, Port-level, and IP-level features are excluded for GQUIC, IQUIC and HTTPS respectively based on UDP and TCP, and our experiment focuses on a single connection between two hosts each time. Selected proven effective features construct the Transfer features, which are specifically defined as:

\begin{itemize}
\item \textbf{unique packet size}: this feature statistics whether the packet $t$ of length $l$ is in the traffic $T$. Specifically, define $\text{Length}(t)$ as a function that calculates the length of packet $t$, if $ l \in \{\text{Length}(t_i)| t_i\in T\}$, $l$-th dimension of this feature is set to 1, otherwise, is set to 0. This feature is a 1460-dimension vector (packet length range from 54 to 1514).
\item \textbf{packet size count}: this feature statistics the number of packet $t$ of length $l$ in the traffic $T$. Specifically, if $ l \in \{\text{Length}(t_i)| t_i\in T\}$, $l$-th dimension of this feature is set to $\text{Card}(\{t_i|t_i\in T, \text{Length}(t_i)=l\})$. This feature is a 1460-dimension vector.
\item \textbf{packet order}: this features records the packets length in order of packet position. Specifically, the $i$-th dimension of this feature is set to $\text{Length}(t_i)$, where $t_i$ is the $i$-th packet in traffic $T$. This feature is a $k$-dimension vector.
\item \textbf{inter-arrival time}: this feature statistics arrival interval of adjacent packets in order of packet position. Specifically, define $\text{Time}(t)$ as a function that fetch the arrival time of a packet, let $t_0$ be the 2-nd \emph{Client Hello} packet for GQUIC, the last Handshake packet for IQUIC, and the \emph{Change Cipher Spec} packet for HTTPS, then $l$-th dimension of this feature is set to $(\text{Time}(t_i)-\text{Time}(t_{i-1}))$, where $ t_i,t_{i-1}\in T$. This feature is a $k$-dimension vector.
\item \textbf{negative packets}: this feature statistics the number of packet $t$ in negative direction in traffic $T$. This 1-dimension feature is set to $\text{Card}(\{(t_j,d_j)| d_j = \text{negative}\})$.
\item \textbf{cumulative size}: this feature statistics the cumulative size of packets in traffic $T$. This 1-dimension feature is set to $\sum \{T_p, T_n\}$, where $T_p = \{\text{Length}((t_i,d_i))| t_i\in T, d_i=\text{positive}\}, T_n = \{\text{Length}((t_i,d_i))| t_i\in T, d_i=\text{negative}\}$.
\item \textbf{cumulative size with direction}: this feature statistics the cumulative size of packets in traffic $T$, but the impact of packet direction $d$ is considered. This 1-dimension feature is set to $\sum \{T_p, T_n\}$, where $T_p = \{\text{Length}((t_i,d_i))| t_i\in T, d_i=\text{positive}\}, T_n = \{-\text{Length}((t_i,d_i))| t_i\in T, d_i=\text{negative}\}$.
\item \textbf{bursts numbers/maximal length/mean length}: burst is define as the consecutive packets between two packets sent in the opposite direction\cite{dyer2012peek}. Bursts numbers, bursts maximal length, and bursts mean length is the statistical features based on burst in the traffic $T$.
\item \textbf{total transmission time}: this feature statistics the total  transmission time of traffic $T$. This 1-dimension feature is set to $\sum \{\text{Time}(t_i)-\text{Time}(t_{i-1})| t_i\in T, i>1\}$.

\end{itemize}

This group of features is defined as Transfer features, proven originally effective in the attack on HTTPS. Transfer features are not conducive to handling high-speed network steam but can adequately expose the vulnerability of protocols, as they are informative but also computationally expensive.

\section{Evaluation}
\label{Evaluation}

In this section, we first introduce the metric and the basic settings used in the experiment. Then, we conduct fair comparisons between GQUIC, IQUIC, and HTTPS and study the vulnerability of protocols to WFP attacks under different scenarios. Furthermore, we analyze the feature importance, feature transferability, and the latent feature representation on different protocol. We also illustrate the high risk of GQUIC and IQUIC by performing upgraded WFP top-$a$ attacks. Finally, we implement WFP attacks on IQUIC under different network conditions, showing that the vulnerability of protocol is only slightly dependent on the network environment. 

\subsection{Evaluation Setup}

\textbf{Metric.} The purpose of the WFP attacks is to infer the website that the user is visiting, which can be seen as a multi-class classification task. In our balanced dataset, accuracy (i.e., attack success rate) is the most intuitive measurement. Higher accuracy means a better attack performance and a more vulnerable protocol. 

\textbf{Model}. In this paper, we do not focus on the improvement of attack algorithms, and five standard machine learning algorithms are utilized as the baseline attack models: \emph{Random Forest (RF), Extra Trees (ET), K-Nearest Neighbors (KNN), Naive Bayes (NB)} and \emph{Support Vector Machine (SVM)}. The algorithms are implemented with scikit-learn\footnote{\url{https://scikit-learn.org}}, using the default parameters. The given experiment results are obtained through 10-fold cross-validation.

\textbf{Scenario}. We took 40 different parameters $k$ from 5 to 200 in steps of 5 to simulate different levels of attack scenario (early $\rightarrow$ normal). We specifically define early traffic scenario as conditions that $k \leq 40$, and normal traffic scenario when $k>40$.

\subsection{Protocols Vulnerability}

\begin{figure}[ht]
\centering
\includegraphics[width=\textwidth]{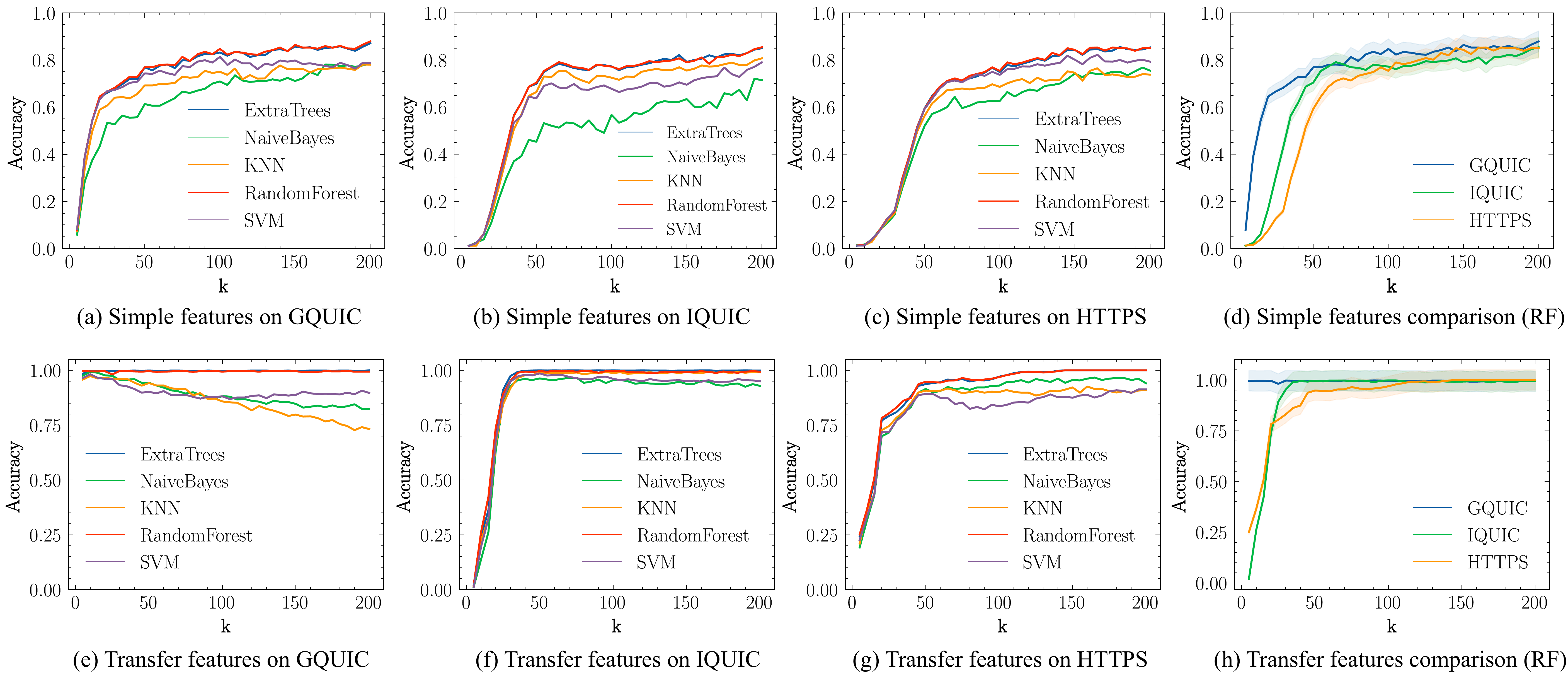}
\caption{Attack accuracy utilizing different features on different scenarios and protocols}
\label{f4}
\end{figure}

Information that can be represented by the features is not saturated in early traffic scenario, and the main factor affecting the attack effectiveness becomes the information richness of the features as $k$ increasing. The attack accuracy increases with parameter $k$ when Simple features are used in the attack (\Cref{f4}(a)-(c)), as a larger $k$ capture more significant differences between websites. When $k$ becomes large enough, the attack accuracy will not continue to increase but will fluctuate steadily over a range, which is also the performance upper bound of a normal WFP attack.

Tree-based models like RF and ET outperformed the five selected attack algorithms on both GQUIC, IQUIC, and HTTPS. When utilizing Simple features, RF and ET respectively reach an average accuracy of 77.3\% and 76.8\% (\Cref{f4}(a)) when on GQUIC, reach 68.5\% and 68.4\% when on IQUIC (\Cref{f4}(b)) , and reach 65.0\% and 64.7\% (\Cref{f4}(c)) when on HTTPS. When using Transfer features, tree-based models also show significant advantages, and their accuracy increases with $k$ as the accuracy of other algorithms slightly decreases when attacking GQUIC and IQUIC. Especially for the attacks on GQUIC, the accuracy of tree-based models is maintained at extremely high level (RF: 99.49\%; ET: 99.78\%, average, \Cref{f4}(e)). RF and ET reach an average accuracy of 94.1\% and 93.2\% when attacking on IQUIC (\Cref{f4}(f)), and reach 91.60\% and 91.30\% (\Cref{f4}(g)) when attacking on HTTPS.

In the early traffic scenario, GQUIC is the most vulnerable to WFP attack among the three protocols, while IQUIC is more vulnerable than HTTPS, but the vulnerability of the three protocols is similar in the normal traffic scenario, when utilizing both Simple features and Transfer features. We find that the accuracy of the attack on three protocols is similar when $k$ is large (e.g., $k=200$, RF, Simple/Transfer, 87.9\%/99.4\% on GQUIC, 85.4\%/99.4\% on IQUIC, 85.1\%/100\% on HTTPS, \Cref{f4}(d)(h)). However, the attack performance shows a large difference when the $k$ is small (e.g., $k=35$, Simple / Transfer, 70.5\%/99.6\% on GQUIC, 53.4\%/98.9\% on IQUIC, 29.4\%/86.2\% on HTTPS, \Cref{f4}(d)(h)). The small $k$ indicates an early traffic scenario, and the gap demonstrates the high security risks of GQUIC and IQUIC when even only a few packets are available. By comparing the result on both simple and complex features, we show that the vulnerability in early traffic scenarios is not caused by the defects of features but the attributes of the protocol itself. Focus on the performance of the attack on GQUIC and IQUIC, and we find that attack can still achieve good performance when only a few packets are considered, even if the features are sample and computationally cheap (Simple feature, $k\leq40$, GQUIC, $\text{accuracy}\leq72.9\%$; IQUIC, $\text{accuracy}\leq61.9\%$; HTTPS, $\text{accuracy}\leq39.7\%$).

\begin{figure}[ht]
\centering
\includegraphics[width=\textwidth]{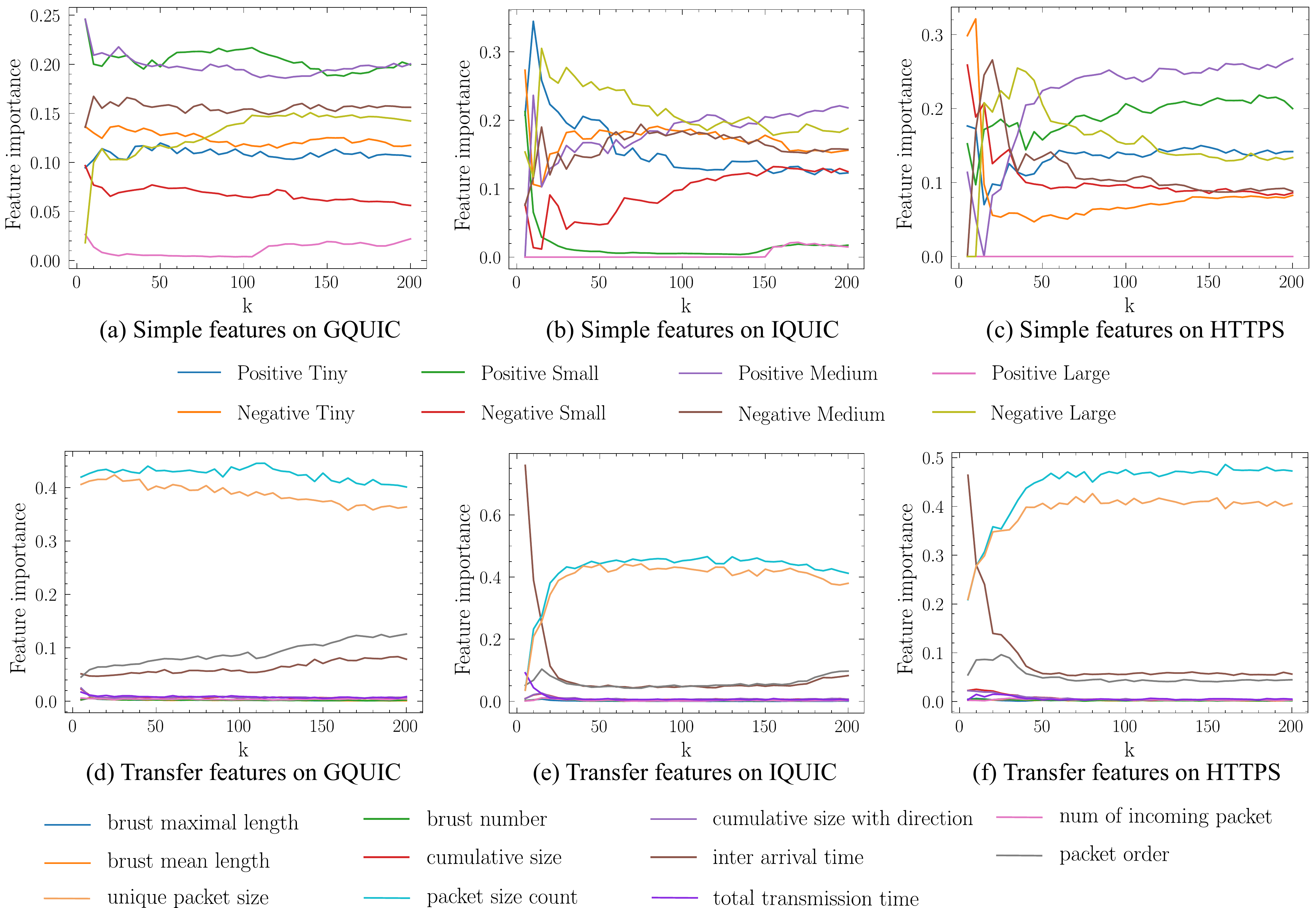}
\caption{Feature importance of Simple features and Transfer features on different scenarios and protocols}
\label{f5}
\end{figure}

\subsection{Feature Importance and Transferability}

In this section, we further explore the feature importance and feature transferability, and expect a new perspective to explain the vulnerability of GQUIC, IQUIC, and HTTPS. All comparisons in our experiments are conducted under the same website distribution, ensuring that the feature importance indicating the effectiveness of the feature on different protocols accurately, which can be evidenced by (i) the upper bound of the attack accuracy and (ii) the number of packets that are needed to achieve high accuracy. We defined \emph{feature transferability} as the effect that a feature is proven effective in the attack on one protocol directly used to attack other protocols, and attribute the transferability to the relatively fixed browser rendering sequence and the similar request-response conversation model. 

Feature importance on IQUIC and HTTPS fluctuate dramatically when $k$ is small, indicating that they will show more different vulnerability when the number of packets considered ($k$) increases, demonstrating that the features considered by the attack algorithm change significantly, and thus indicates that the traffic distribution of the protocol changes significantly, which directly leads to the difficulty of attacking the protocol. However, the drastic change region of features importance on IQUIC is more concentrated ($k<20$), while is more diffuse on HTTPS ($k<50$), indicating that traffic distribution of IQUIC will reach a steady-state earlier, thus exposing vulnerability to WFP attacks earlier, resulting in that fewer packets are needed to achieve the same attack efficiency as when attacking HTTPS  (e.g., Transfer features, $k=30$: IQUIC, 97.4\%; HTTPS, 83.0\%), as high attack efficiency is associated with slight feature importance fluctuation. The feature importance is always more stable when the attack implements on GQUIC than on HTTPS (Simple features: GQUIC, $\sigma ^2 = 1.275E-04$; HTTPS, $\sigma ^2 = 1.736E-03$. Transfer feature: GQUIC, $\sigma ^2 = 9.940E-05$; HTTPS, $\sigma ^2 = 1.081E-03$), indicating that the traffic distribution of GQUIC is not associated with the number of packets considered, and even few packets can expose the website information hidden behind (e.g., early traffic, average, GQUIC: 55.34\%; IQUIC: 27.18\%; HTTPS: 13.98\%).

\begin{figure}[htbp]
\centering
\includegraphics[width=\textwidth]{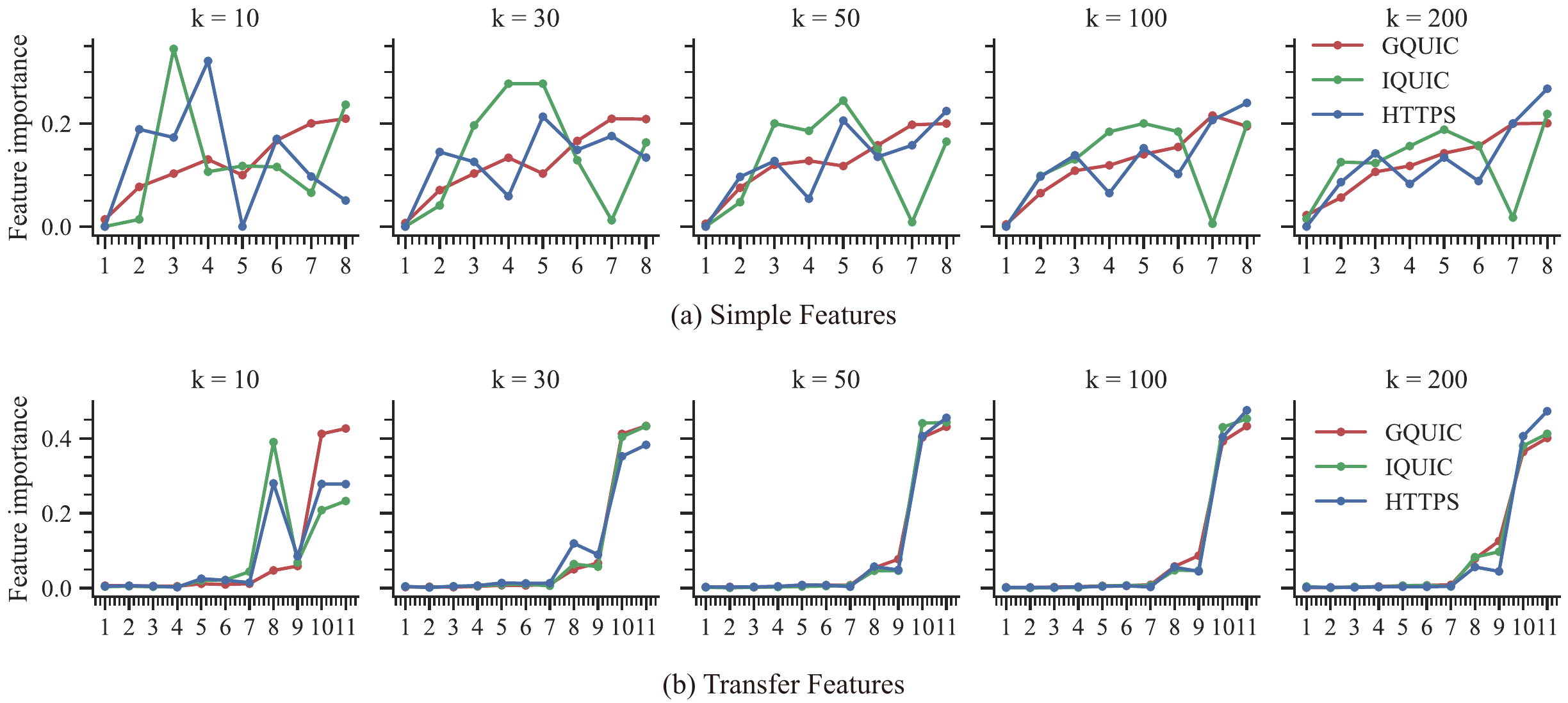}
\caption{Inheritance of feature importance on different scenarios and protocols}
\label{f7}
\end{figure}

To illustrate the inheritance of feature across GQUIC, IQUIC, and HTTPS, we rank the features on GQUIC by feature importance at $k=200$ and \Cref{f7} shows the feature importance on GQUIC, IQUIC, and HTTPS in the fixed order under different $k$. We find that the feature importance changes proportionally over the three protocols when $k\geq30$, except that the Positive Small in the Simple features is always inefficiency when attack IQUIC, as client few to send small packets to server in the normal traffic when $k<200$. However, the inheritance is ineffective in the early traffic scenario, as the distribution of IQUIC and HTTPS traffic are much more variable than GQUIC, and the drastic change region of features importance on IQUIC is more concentrated, while is more diffuse on HTTPS, resulting in that the difference in features importance on the three protocols is more extensive when the value of $k$ is small. Specifically, when on normal traffic scenario, \emph{Positive Medium} and \emph{Negative Large} in Simple features are effective in the attack on both GQUIC, IQUIC, and HTTPS, and \emph{Positive Large} always contribute a little (\Cref{f5}(a)-(c)). Similarly, \emph{packets size count, unique packets size, inter-arrival time}, and \emph{packet order} are significant in both attacks, ranking top 4 among Transfer features while the importance of other features are close to zero (\Cref{f5}(d)-(f)).

\subsection{Qualitative Analysis of Latent Representation}
\begin{figure}[ht]
\centering
\includegraphics[width=\textwidth]{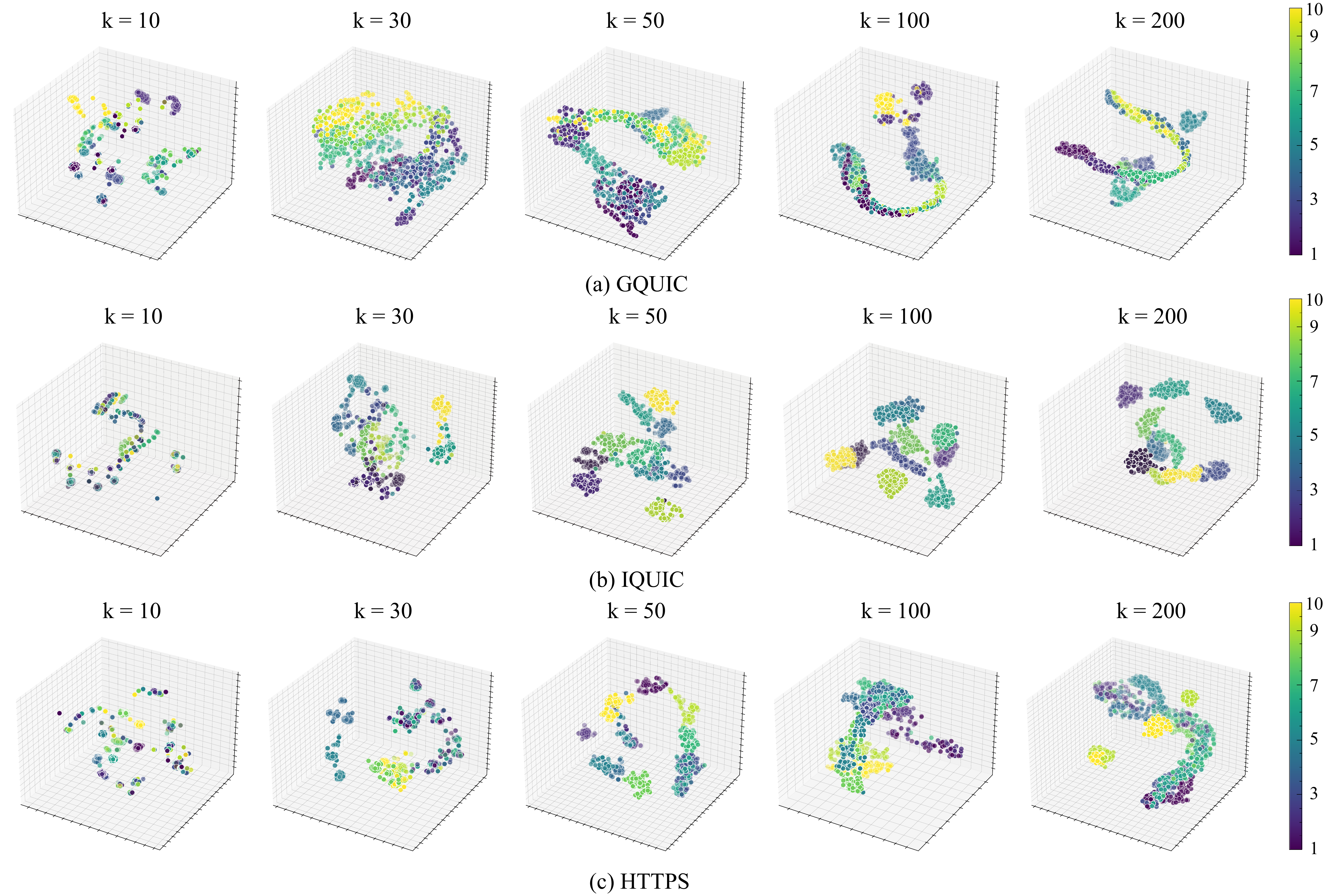}
\caption{Latent representation of Transfer features on different scenarios and protocols}
\label{f6}
\end{figure}

In this section, we qualitatively analyze the vulnerability differences in terms of the features latent representation on the traffic of GQUIC, IQUIC, and HTTPS. The qualitative analysis, combined with the quantitative analysis in the previous section, will provide a more comprehensive description of the nature of the protocol vulnerability. To better display the distribution pattern in latent space, we randomly select 10 out of the 92 websites and analyze only the sampled data. We also select five $k$ values with unequal differences to better represent the process of variation as the degree of latent representation variation of the feature over the three protocols will be more considerable when $k$ is small and will gradually stabilize as $k$ increases. We utilize the Transfer features as the original representation as it exposes the vulnerable characteristic of protocol more completely. T-distributed Stochastic Neighbor Embedding (T-SNE) is used to extract latent representation of the protocol on the sampled data at fixed $k$ values (\Cref{f6}).

When $k=10$, i.e., in the extremely early traffic scenario, the latent features representation of both GQUIC, IQUIC and HTTPS have many samples collapse together, which indicates that different classes of samples are difficult to be represented clearly, thus illustrating the difficulty of conducting WFP attacks on early traffic. However, even in this case, the representation space of GQUIC is more of aggregations of the same class samples, while the representation of IQUIC and HTTPS is more of aggregations of the different class samples. This indicates that GQUIC, in the extremely early scenario, although the intra-class gap cannot be fully learned yet, the adequate representation of inter-class gap has laid the groundwork for the high attack success rate. When $k=30$, as the increased number of considered packets, the Transfer features can better express the inter-class disparity. There is a clear tendency of dispersion in the feature representation on GQUIC and IQUIC, especially GQUIC, and all samples are more spreading, which indicates that the latent representation space of GQUIC and IQUIC can distinguish the inter-class data better, making the protocol more vulnerable to attacks. The representation on HTTPS also shows the same trend, but the magnitude of change is relatively small, and most of the samples are still too aggregated to be represented, which illustrates the defensive of HTTPS against WFP attacks. When $k=50$, the representation space of GQUIC and IQUIC have almost convergence, and when $k$ continues to increase, the relative positions between points in the space are very similar, which mirrors the results obtained when the Transfer feature and RF are used to attack GQUIC and IQUIC when $k>50$. It should be noted that there is still a large mixture of samples of different classes in the HTTPS representation space. When $k>100$, the representation space of GQUIC, IQUIC and HTTPS show similar characteristics, and the similarities and differences between intra-class samples and inter-class samples are well represented, indicating that the vulnerabilities of GQUIC, IQUIC and HTTPS are similar under the normal traffic scenario, which is consistent with the previous results of the quantitative analysis.

\subsection{Protocols vulnerability under Top-a Attack}

In this section, we expect to illustrate the high risk of GQUIC and IQUIC when they suffer from multiple WFP Top-$a$ attacks, which means the adversary will predict $a$ websites with the highest probability as the target website. As recall is more concerned by the adversary than precision, the adversary will perform further attacks if the first attack fails to infer the target website. Specifically, Top-$a$ attack is defined as:

\begin{equation}
M(F,\theta) \longrightarrow A
\label{WFPTopattack}
\end{equation}

\begin{equation}
M(\text{Early}(F, k),\theta) \longrightarrow A
\label{EWFPTopattack}
\end{equation}

Where, $A=\cup_{i=1}^{a}\{\{W_i| i \in \{1,2,...,N\}, P(W_1)>P(W_2)>...>P(W_N)\}\}$, $P(W_i)$ is the vising probability of website $W_i$ given by the model. A Top-$a$ attack is counted as success when the website accessing by the user belongs to the prediction set, i.e., $W \in A$. It should be noted that upgrading from a WFP attack to a Top-$a$ attack does not require additional calculations, ensuring the efficiency does not drop, even dealing with a high-speed stream.

The adversary (or the secure gateway) should prefer to use computationally cheap features during the attack to reduce the latency of the attack, thus saving more time to prepare for further attacks. Moreover, since the attack's accuracy on GQUIC fail to indicate the difference between protocols under Top-$a$ attack when using Transfer features and RF (accuracy are high enough due to the complexity of features, i.e., always close to 100\% when on GQUIC), we conduct the comparisons utilizing simple and effective Simple features while using the most effective RF as the attack model.

\input{tables/table2}

When the attack upgrade to a Top-$a$ attack, the weaknesses of GQUIC and IQUIC are further exposed in two ways (\Cref{t2}). First, Top-$a$ attacks benefit more when implemented on GQUIC and IQUIC than on HTTPS. As the parameter $a$ increases ($1\rightarrow5$), and the attack accuracy improves more on GQUIC and IQUIC (average, GQUIC: 27.0\%; IQUIC: 26.1\%; HTTPS: 14.7\%). At the most significantly improve scenario, the improvement on GQUIC and IQUIC is almost two times as the improvement on HTTPS (GQUIC: 41.2\%; IQUIC: 40.8\%; HTTPS: 22.9\%). This attack effectiveness improvement is independent of the attack accuracy of baseline, i.e., accuracy when $a=1$, but is determined by the characteristic of GQUIC and IQUIC. Even attacks has similar accuracy on both GQUIC, IQUIC and HTTPS when $a=1$, the Top-$a$ attack on GQUIC and IQUIC always achieves a higher accuracy improvement (e.g., GQUIC: k=10, accuracy=38.7\%, improve: 41.2\%; IQUIC: $k=30$, accuracy=42.8\%, improve:40.8\%; HTTPS: k=40, accuracy=39.7\%, improve: 21.0\%). This vulnerability of GQUIC and IQUIC makes them especially 
insecure that an adversary can always achieve a significant accuracy improvement within an acceptable tolerance even when the attack accuracy is originally low. Second, Top-$a$ attacks can achieve high accuracy on GQUIC and IQUIC, especially GQUIC, with a tiny number of packets. The accuracy gap on GQUIC reaches the largest when using only 10 packets, and the Top-5 attack can identify the website currently being transmitted over GQUIC with an accuracy of 79.9\%, while the accuracy of the attack on HTTPS is only 7.2\% under the same conditions. The most significant improvement on IQUIC happens when $k=30$, and the accuracy of Top-5 attack on IQUIC achieves 83.6\%, while only 38.7\% when on HTTPS. Moreover, when the attack reaches the best performance, using only 40 packets, attack accuracy could reach 95.4\% when on GQUIC,  reach 95.5\% when on IQUIC, whereas only 60.7\% when on HTTPS.

\subsection{Vulnerability Under Flawed Network}

Network imperfect factors can affect the interaction between client and server, e.g., high loss can lead to more retransmission in traffic, which affects the effectiveness of features on traffic representation and thus the efficiency of WFP attack. In this section, to illustrate that the vulnerability of the protocol is only slightly dependent on the network conditions, we implement the WFP attacks on IQUIC in different network conditions. Specifically, we control the bandwidth, delay, and loss of the network and collect the traffic generated when user visit 50 randomly selected websites. We first conduct the Top-$a$ attacks utilizing Simple features to show that the adversary can perform effective attacks even only utilize the computationally cheap simple feature under flawed networks. We also conduct WFP attacks utilizing Transfer features to demonstrate that informative features can effectively represent different traffic even under flawed networks.

\begin{figure}[ht]
\centering
\includegraphics[width=\textwidth]{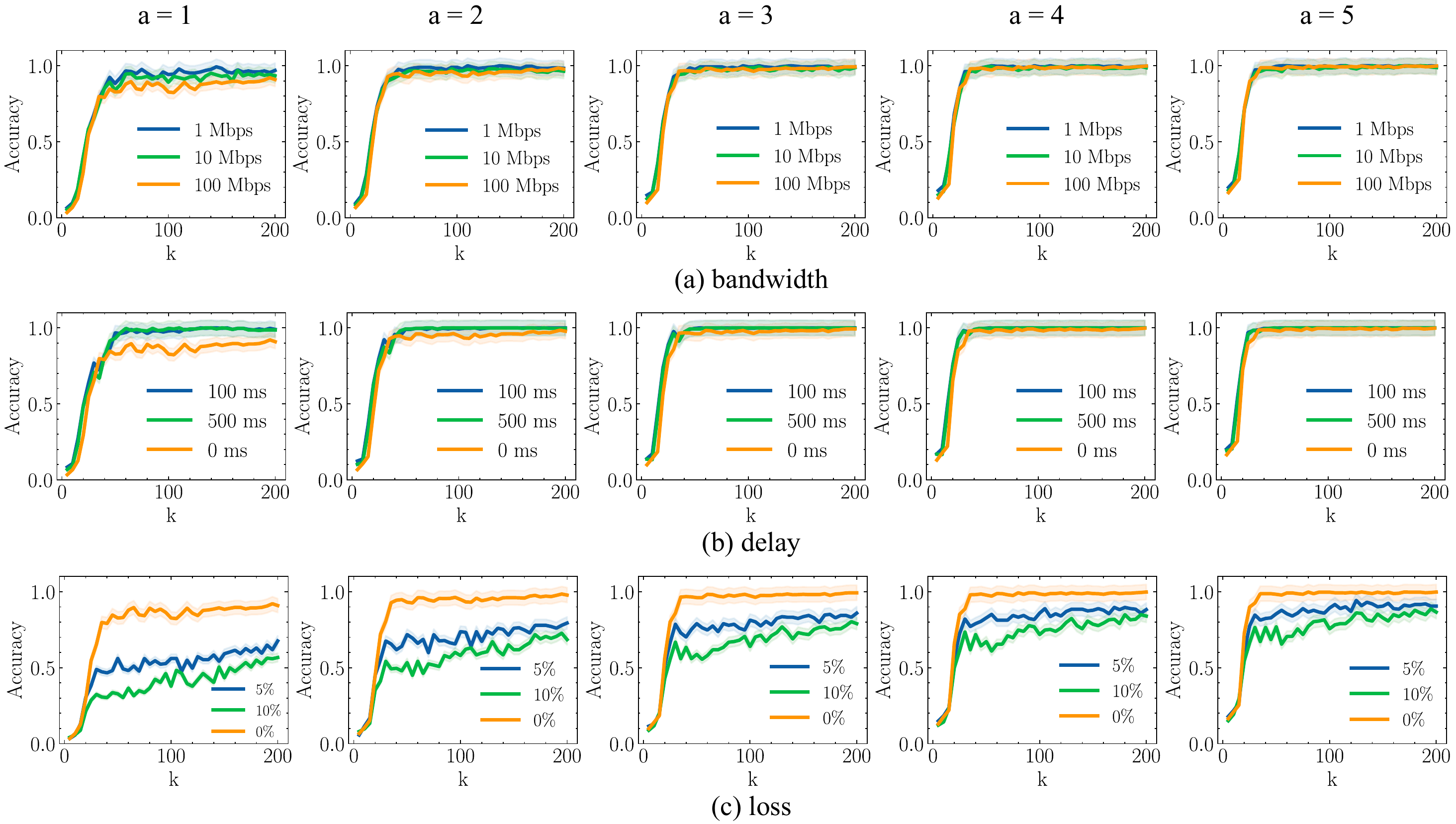}
\caption{Attack accuracy on IQUIC utilizing Simple features under different network conditions and scenarios}
\label{f8}
\end{figure}

\begin{figure}[ht]
\centering
\includegraphics[width=\textwidth]{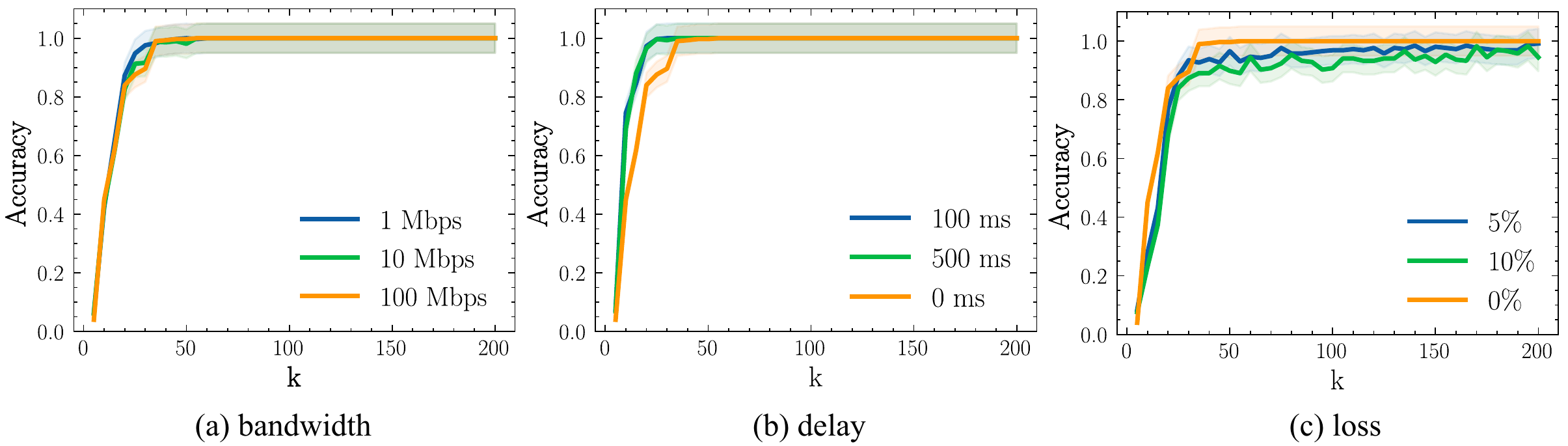}
\caption{Attack accuracy on IQUIC utilizing Transfer features under different network conditions and scenarios}
\label{f9}
\end{figure}

The bandwidth and delay have no negative impact on the effectiveness of the attack when both utilizing Simple features and Transfer features. A slight decrease in bandwidth and increase in delay will make the features more fully represent the traffic, thus improving the effectiveness of WFP attacks (\Cref{f8}(a)(b), \Cref{f9}(a)(b)). When utilizing Simple features, the effectiveness of WFP attacks in normal traffic scenario is improved at low bandwidth and high delay ($a=1, k=200$, bandwidth, 1 Mbps: 96.9\%, 10 Mbps: 93.4\%, 100 Mbps: 90.9\%; delay, 500 ms: 98.6\%, 100 ms: 99.0\%, 0 ms: 90.9\%). But in the early traffic scenario, bandwidth and delay have little impact on the effectiveness of the attack ($k=30$, bandwidth, 1 Mbps: 67.4\%, 10 Mbps: 64.9\%, 100 Mpbs: 65.3\%; $k=40$, delay, 500 ms: 81.7\%, 100 ms: 82.3\%, 0 ms: 79.1\%). When utilizing Transfer features, the effectiveness of WFP attacks is almost the same under different bandwidth and delay condition (average, bandwidth, 1 Mbps: 94.2\%, 10 Mbps: 94.8\%, 100 Mbps: 94.3\%;  delay, 500 ms: 96.5\%, 100 ms: 96.9\%, 0 ms: 94.3\%).

High loss make the WFP attacks more difficult (\Cref{f8}(c), \Cref{f9}(c)), resulting in the accuracy decrease of the attack under both early traffic scenario and normal traffic scenario, especially when utilizing Simple features (Simple features, $k=40$, loss 0\%: 79.1\%, loss 5\%: 46.6\%, loss 10\%: 30.3\%; $k=200$, loss 0\%: 90.9\%, loss 5\%: 67.6\%, loss 10\%: 56.7\%). Moreover, when utilizing Simple features, the high loss also makes the inflection point of the attack accuracy from rapidly increasing to steady appear earlier (loss 0\%: $k=35$; loss 5\%: $k=30$, loss 10\%: $k=25$), which further makes the WFP attack in the early traffic scenario difficult. We believe that the effectiveness of Simple features decreases in the high loss condition is due to the large number of retransmission, which introduces uncertainty to the relatively fixed sequence of packets originally caused by the resource structure of the website and the resource access paradigm of browsers under the request-response conversation model between client and server. The distribution of samples is shifted in an uncertain direction, and the space of different classes is crossed, making it difficult to represent the samples of different classes. However, the negative impact of packets loss is significantly reduced when attacks utilizing Transfer features, and the attack accuracy gap under different loss rates are much smaller than when utilizing Simple features (average, loss, Transfer feature: 0\%: 94.2\%, 5\%: 90.6\%, 10\%: 87.1\%; Simple feature: 0\%: 78.3\%, 5\%: 50.7\%, 10\%: 35.9\%). This indicate that informative features can better resist the negative impact of flawed network. It should be noted that even under a poor network, as the loss reaches 5\% or 10\%, the trend of the attack accuracy with the increase of parameter $k$ remains similar, which increases sharply in the early traffic scenario and then tends to be steady. Therefore, we believe that the trend of the attack accuracy on GQUIC and HTTPS will also remain similar to which under perfect network, indicating that the relative vulnerability of GQUIC, IQUIC, and HTTPS to WFP attacks on early traffic is only slightly affected by the network conditions.

More importantly, as the parameter $a$ increases, the drop of accuracy caused by the network imperfection decreases, indicating that even in a poor network condition, the adversary can still make the attack accuracy close to the result in a perfect network by increasing the tolerance of the attack, i.e., increasing the number of attacks ($a=5, k=200$, loss, 0\%: 99.8\%, 5\%: 90.5\%, 10\%: 86.6\%), which also shows that the effect of the network condition is slight for the adversary, and the adversary can always find a new balance between the attack efficiency and effectiveness by increasing the complexity of the features.

\section{Discussion}
\label{Discussion}

The superior transmission performance of GQUIC and IQUIC brings opportunities for speeding up the Internet, but protocols security risks also bring uncertainties. Previous researches \cite{tong2018novel,rezaei2020multitask} explored the website fingerprinting of QUIC on a normal traffic scenario, while we focus more on the risk of different versions of QUIC on early traffic. The vulnerability of GQUIC and IQUIC on early traffic poses a significant challenge to the privacy and confidentiality guaranteed.

One limitation of our work is that we do not conduct the experiments under a real open world, where the network condition may be much more complicated than the local area network. However, conducting experiments in a controlled environment also allows us to have more precise control over the network environment, and thus can explore the impact of different network factors on protocol vulnerability, which is difficult to achieve precisely in an open network. Moreover, our testbed provides a realistic simulation of real situations, and our research focuses on the vulnerability of GQUIC, IQUIC, and HTTPS under the same condition, rather than the attack performance against a single protocol, we believe that the conclusions we have achieved can generalize to a real open network. In the future, we plan to explore how features are transferred and applied across different protocols, which may help us to construct generalizable attack models.

\section{Conclusion}
\label{Conclusion}

Cryptographic protocol can protect the user's privacy and avoid exposing private information to the adversary. In this paper, we discuss the vulnerability of GQUIC, IQUIC, and HTTPS to WFP attacks and the feasibility of feature transferring between protocols on both early traffic and normal traffic scenarios. We demonstrate that GQUIC and IQUIC, especially GQUIC, are more vulnerable to WFP attacks than HTTPS in the early traffic scenario but are similar in the normal traffic scenario. We also demonstrate that, when on normal traffic, most features are transferable between protocols, and the feature importance is inherited. However, the transferring is inefficient when on early traffic due to the different magnitude of variation in the traffic distribution of different protocols. Moreover, we quantitatively analyze the latent feature representation space of GQUIC, IQUIC, and HTTPS to intuitively show that features can represent inter-class and intra-class samples more efficiently on GQUIC and IQUIC than on HTTPS, causing GQUIC and IQUIC more vulnerable when on early traffic scenario. We also show that an adversary can always achieve a significant attack efficiency improvement within an acceptable tolerance even when the attack accuracy is originally low on GQUIC and IQUIC. Finally, we conduct experiments on different flawed networks to show that the vulnerability of GQUIC, IQUIC, and HTTPS is slightly dependent on the network environment, and our conclusion can generalize to a real open network.

\section*{Acknowledgments}
This work was supported by National Key R\&D Program of China under grant NO.2018YFB0803604. The authors would like to thank the editor and anonymous reviewers for their comprehensive and constructive comments.


\bibliographystyle{elsarticle-num} 
\bibliography{main.bib}

\end{document}

%% file: tables/table2.tex
\begin{table}[htbp!]
\centering
\caption{Accuracy of Top-$a$ attacks with Random Forest on different scenarios and protocols}
\label{t2}
\begin{tabular}{cc|rrrrr|c}
\toprule
&  & \multicolumn{5}{c|}{$a$} & \multicolumn{1}{c}{improve} \\ \cmidrule(l){3-7} 
                       \multicolumn{1}{c}{Protocol}&  \multicolumn{1}{c|}{$k$}  & \multicolumn{1}{c}{1}     & \multicolumn{1}{c}{2}     & \multicolumn{1}{c}{3}     &  \multicolumn{1}{c}{4}     & \multicolumn{1}{c|}{5}              &    \multicolumn{1}{c}{\footnotesize{($1\rightarrow5$)}}            \\ \midrule
\multicolumn{1}{c|}{\multirow{8}{*}{GQUIC}}  & 5  & 7.9 & 14.0  & 18.3 & 22.0  & 26.0           & 18.1          \\
                      \multicolumn{1}{c|}{} & 10 & 38.7 & 58.3 & 68.2 & 75.7 & 79.9          & $\mathbf{41.2}$ \\
                      \multicolumn{1}{c|}{} & 15 & 54.2 & 70.2 & 78.6 & 83.9 & 87.2          & 33.0           \\
                      \multicolumn{1}{c|}{} & 20 & 64.5 & 78.8 & 85.2 & 88.7 & 91.1          & 26.6          \\
                      \multicolumn{1}{c|}{} & 25 & 66.6 & 81.0  & 87.5 & 91.2 & 93.0           & 26.4          \\
                      \multicolumn{1}{c|}{} & 30 & 68.2 & 80.6 & 87.1 & 90.6 & 92.7          & 24.5          \\
                      \multicolumn{1}{c|}{} & 35 & 70.5 & 83.8 & 88.8 & 92.5 & 94.1          & 23.6          \\
                      \multicolumn{1}{c|}{} & 40 & 72.9 & 85.2 & 91.0  & 94.3 & $\mathbf{95.4}$ & 22.5          \\ \midrule
\multicolumn{1}{c|}{\multirow{8}{*}{IQUIC}} & 5  & 1.1 & 2.3 & 3.5 & 4.6 & 5.7 & \hspace{0.5em}4.6          \\
                      \multicolumn{1}{c|}{} & 10 & 2.4 & 4.6 & 6.1 & 7.3 & 8.3 & \hspace{0.5em}5.9          \\
                      \multicolumn{1}{c|}{} & 15 & 6.1 & 11.5 & 15.3 & 18.3 & 20.5 & 14.4          \\
                      \multicolumn{1}{c|}{} & 20 & 16.8 & 26.5 & 35.1 & 42.3 & 49.1 & 32.3          \\
                      \multicolumn{1}{c|}{} & 25 & 30.1 & 45.6 & 55.2 & 63.1 & 69.8 & 39.7          \\
                      \multicolumn{1}{c|}{} & 30 & 42.8 & 60.0 & 71.4 & 78.7 & 83.6 & $\mathbf{40.8}$ \\
                      \multicolumn{1}{c|}{} & 35 & 56.3 & 73.1 & 83.6 & 91.1 & 93.4 & 37.1          \\
                      \multicolumn{1}{c|}{} & 40 & 61.9 & 81.4 & 89.1 & 93.2 & $\mathbf{95.5}$ & 33.6           \\ \midrule
\multicolumn{1}{c|}{\multirow{8}{*}{HTTPS}} & 5  & 1.3 & 2.4 & 3.9 & 5.3 & 6.8          & \hspace{0.5em}5.5          \\
                      \multicolumn{1}{c|}{} & 10 & 1.5 & 3.1 & 4.3 & 5.8 & 7.2          & \hspace{0.5em}5.7          \\
                      \multicolumn{1}{c|}{} & 15 & 3.8 & 7.4 & 9.6 & 12.1 & 14.5          & 10.7          \\
                      \multicolumn{1}{c|}{} & 20 & 7.7 & 11.2 & 14.9 & 18.2 & 21.9          & 14.2          \\
                      \multicolumn{1}{c|}{} & 25 & 12.6 & 18.6 & 22.4 & 25.9 & 28.9          & 16.3          \\
                      \multicolumn{1}{c|}{} & 30 & 15.8 & 22.9 & 29.6 & 34.2 & 38.7          & $\mathbf{22.9}$ \\
                      \multicolumn{1}{c|}{} & 35 & 29.4 & 37.5 & 43.4 & 47.7 & 50.7          & 21.3          \\
                      \multicolumn{1}{c|}{} & 40 & 39.7 & 50.1 & 55.8 & 58.1 & $\mathbf{60.7}$ & 21.0           \\ \bottomrule
\end{tabular}
\end{table}